\pdfoutput=1
\documentclass[11pt,a4paper]{article}
\usepackage{jcappub}
\allowdisplaybreaks
\usepackage{xcolor}
\usepackage{amsmath}
\usepackage{graphicx}
\usepackage{esint}
\usepackage{slashed}
\usepackage{bigints}
\usepackage[english]{babel}
\usepackage{comment}
\usepackage{hyperref}

\usepackage{babel}
\begin{document}

\title{Revisiting supernova constraints on a light CP-even scalar}

\author[a]{P. S. Bhupal Dev,}
\author[b]{Rabindra N. Mohapatra,}
\author[a]{Yongchao Zhang}
\affiliation[a]{Department of Physics and McDonnell Center for the Space Sciences,  Washington University,
St. Louis, MO 63130, USA}
\affiliation[b]{Maryland Center for Fundamental Physics, Department of Physics, University of Maryland,
College Park, MD 20742, USA}
\emailAdd{bdev@wustl.edu, rmohapat@umd.edu, yongchao.zhang@physics.wustl.edu}

\abstract{ A light CP-even Standard Model (SM) gauge-singlet scalar $S$ can be produced abundantly in the supernova core, via the nucleon bremsstrahlung process $N N \to N N S$,  due to its mixing with the SM Higgs boson. Including the effective $S$ coupling to both nucleons and the pion mediators, we evaluate the production amplitude for the $S$ particle and point out a key difference with the well-known light CP-odd scalar (axion) and vector boson (dark photon) cases. Taking the subsequent decay and re-absorption of $S$ into account, we present a complete calculation of the energy loss rate for the $S$ particle. We then use the SN1987A luminosity constraints to derive updated supernova limits on the mixing of the scalar $S$ with the SM Higgs boson. We find that the mixing angle $\sin\theta$ with the SM Higgs is excluded only in the narrow range of $3.9 \times 10^{-7}$ to $7.0 \times 10^{-6}$, depending on the scalar mass up to about 147 MeV, beyond which the supernova limit disappears.  }
%\blue{\sout{This result has important consequences for the laboratory searches for light scalars.}}

\keywords{Supernova, Beyond Standard Model}

\maketitle
%\tableofcontents

\section{Introduction}

In the absence of any new TeV-scale particles appearing at the Large Hadron collider, the possibility that new physics beyond the Standard Model (SM) of particle physics could manifest itself in the low-energy regime,  e.g. below GeV-scale has recently attracted considerable attention. Typical candidates include light CP-odd scalar particles,  such as the widely discussed QCD axion (originally proposed to solve the strong CP problem~\cite{Peccei:1977hh, Weinberg:1977ma, Wilczek:1977pj}) or similar axion-like particles (ALPs), SM-singlet CP-even scalars, and light vector bosons such as dark photons, all of which can be searched for in the high-intensity experiments~\cite{Jaeckel:2010ni, Hewett:2012ns, Essig:2013lka, Graham:2015ouw, Alexander:2016aln}. Such particles have the important property that if their masses are below a few hundred MeV or so, they can also manifest themselves in the compact astrophysical objects like supernovae or neutron stars and affect their  properties, thereby making these astrophysical systems an ideal laboratory for beyond SM physics search~\cite{Raffelt:1996wa}. %A natural CP-odd scalar is the axion, which is proposed to solve the strong CP problem~\cite{Peccei:1977hh}. Through the couplings to nucleons $N$\footnote{At leading order the axion/ALP production process in the supernova core is mediated by a pion. As pion and axion/ALP are both pseudo-Nambu-Goldstone boson, in calculation of the emission rate the couplings of pion and axion/ALP to nucleons can not be written simultaneously in the form of pseudoscalar, i.e. $N i \gamma_5 N$, without introducing contact terms~\cite{Brinkmann:1988vi}.}
%\begin{eqnarray}
%\label{eqn:ALP}
%{\cal L}_{\rm int} \ \supset \ y (\partial_\mu a) \overline{N} i \gamma_\mu \gamma_5 N \,,
%\end{eqnarray}
For example, the light axion or ALPs ($a$) can be produced abundantly in the supernova core via the nuclear bremsstrahlung process $N N \to N N a$ (where $N=p,n$ collectively stands for protons and neutrons). If their mean free path is larger than the size of the supernova core, these light particles could carry away a significant fraction of the energy. Therefore, their couplings to matter %in Eq.~(\ref{eqn:ALP})
can be constrained~\cite{Iwamoto:1984ir, Pantziris:1986dc, Turner:1987by, Raffelt:1987yt, Mayle:1987as, Brinkmann:1988vi,  Burrows:1988ah} from the inferred energy loss rate ${\cal L}_\nu \sim (3-5) \times 10^{53}$ erg/sec, derived from the detection of late-time neutrino events from SN1987A by Kamiokande~\cite{Hirata:1987hu}. This is the so-called `Raffelt criterion'~\cite{Raffelt:1996wa} (see Ref.~\cite{Bar:2019ifz} for a critical assessment of the Raffelt criterion).

Following the same calculation as in the axion case, supernova limits on a light CP-even scalar ($S$) have been discussed in Refs.~\cite{Ishizuka:1989ts, Diener:2013xpa, Krnjaic:2015mbs, Lee:2018lcj} (see also Ref.~\cite{Arndt:2002yg} for the limit on saxion which to some extent is similar to the $S$ particle). In  this paper, we revisit the calculation for $S$ production and its mean free path in a supernova core, and point out some important differences with the pseudoscalar axion/ALP case, which lead to some modifications of the supernova constraints on $S$, as compared to the existing bounds in the literature.
%supernova luminosity limit on a generic light CP-even scalar $S$,
%as copious production of $S$,
%whose only coupling to SM particles is through its mixing with the SM Higgs boson.
%and study how its production in the supernova core leads to extra energy loss.

The light scalar $S$ has been discussed as a natural dark matter candidate~\cite{Silveira:1985rk, McDonald:1993ex, Burgess:2000yq, Cline:2013gha, Athron:2017kgt} or dark force mediator~\cite{Pospelov:2007mp, Kainulainen:2015sva, Bell:2016ekl, Knapen:2017xzo, Matsumoto:2018acr, Batell:2018fqo}, or a particle which can assist in the generation of the baryon asymmetry of the Universe~\cite{Espinosa:1993bs, Profumo:2007wc, Espinosa:2011ax, Croon:2019ugf}. The singlet scalar $S$ with mass around $100$ MeV and mixing angle with the SM Higgs of $\theta\sim 3\times 10^{-4}$ could also be responsible for the recent anomalous excess in the flavor-changing kaon decay $K_L \to \pi^0 \nu\bar{\nu}$ observed in the KOTO experiment~\cite{Kitahara:2019lws, Egana-Ugrinovic:2019wzj, Dev:2019hho, Liu:2020qgx, Cline:2020mdt}.  This makes a careful (re)examination of the supernova constraints on light scalar even more relevant and timely.
%It is therefore important to study constraints on its couplings from various observations.
%The supernova limit on $S$ has been studied in Ref.~\cite{Ishizuka:1989ts, Diener:2013xpa, Krnjaic:2015mbs, Lee:2018lcj} (see also Ref.~\cite{Arndt:2002yg} for the limit on saxion which is to some extent similar to the $S$ particle, although very different in details). In  this paper we revisit the supernova luminosity limit on a generic light CP-even scalar $S$,
%as copious production of $S$,
%whose only coupling to SM particles is through its mixing with the SM Higgs boson.
%and study how its production in the supernova core leads to extra energy loss.

The couplings of light $S$ to the SM particles can be induced in two ways: (i) through mixing with the SM Higgs $h$, and (ii) through the introduction of new heavy particles such that $S$ can couple radiatively to photons or gluons (or other SM particles) via the heavy particle loops.  A good example of type (ii) couplings naturally occur in the left-right symmetric model (LRSM) based on the gauge group $SU(2)_L \times SU(2)_R \times U(1)_{B-L}$~\cite{Pati:1974yy, Mohapatra:1974gc, Senjanovic:1975rk}, where
%provides a natural candidate of the second type in some regions of parameter space, where
the light scalar originates from the $SU(2)_R$-breaking triplet scalar field, and the heavy $W_R$ boson and heavy scalars in the LRSM mediate the decay $S \to \gamma\gamma$ even if $S$ does not mix with the SM Higgs~\cite{Nemevsek:2012cd, Dev:2016vle, Dev:2017dui}. This restricts the applicability of the supernova bound to only a narrow mass range of $m_S=20$--30 MeV~\cite{Dev:2019hho}. Therefore, we will consider only the $S$ couplings of type (i), where all the couplings of $S$ to the SM particles are induced from the  $h-S$ mixing, with a universal mixing angle $\sin\theta$. Then we can use the Raffelt criterion on energy loss from SN1987A to set limits on the scalar mass $m_S$ and the mixing angle $\sin\theta$.

Through mixing with the SM Higgs, the light scalar $S$ couples to nucleons, pions, electrons and photons. Similar to the axion/ALP case, $S$ can be produced in the supernova core via the following processes~\cite{Pantziris:1986dc}:
\begin{itemize}
    \item nucleon bremsstrahlung $N N \to N N S$;
    \item $e^+ e^-$ annihilation $e^+ e^- \to \gamma S$;
    \item $\gamma\gamma$ annihilation $\gamma\gamma \to S$;
    \item Compton-like scattering $e\gamma \to e S$;
    \item Primakoff-like process $\gamma (e,N) \to S (e,N)$;
    \item plasma ($\gamma_{\rm pl}$) decay $\gamma_{\rm pl} \to \gamma S$.
\end{itemize}
The coupling of $S$ to electrons is highly suppressed by the small Yukawa coupling $y_e$ of the SM Higgs field and the coupling  to photons is loop suppressed. Thus the production of $S$ in the supernova core is dominated by the nucleon bremsstrahlung process.

After being produced in the supernova core, the scalar $S$ can decay into light SM particles such as $\gamma\gamma$, $e^+ e^-$ and $\mu^+ \mu^-$. If these decays happen inside the supernova core, then the energy loss constraints will not apply. The $S$ particle can also scatter with nucleons and be re-absorbed via the inverse nucleon bremsstrahlung process $NNS\to NN$.
%\begin{eqnarray}
%\label{eqn:3to2}
%N + N + S \to N + N \,.
%\end{eqnarray}
Here again we neglect the scattering of $S$ with electrons and photons; for instance, the processes $S e \to e\gamma$ and $S\gamma \to e^+ e^-$ are suppressed by either the small couplings to electrons or by the loop factor from photon coupling. From the scattering rate of the $3\to 2$ re-absorption process we can obtain the mean free path (MFP) of $S$, and compare it to the supernova core radius to see whether the $S$ particle can escape from the core for a given mixing angle. It turns out that for $m_S \lesssim 100$ MeV, the MFP of $S$  is of the order of the supernova core size (roughly 10 km) for a mixing angle of $\sin\theta \sim 10^{-6}$ (see Fig.~\ref{fig:mfp}). After taking into account both the decay and re-absorption of $S$, we find that the supernova luminosity limit excludes the scalar mass up to roughly 147 MeV for mixing angle $\sin\theta$ ranges from $3.9 \times 10^{-7}$ to $7.0 \times 10^{-6}$ (see Fig.~\ref{fig:limit}). % depending on the scalar mass, as seen in Fig.~\ref{fig:limit}.
%Beyond $m_S=2m_\pi$, the channel $S \to \pi\pi$ is kinematically open and the scalar $S$ decays too fast inside the core itself.
Our final result shown in Fig.~\ref{fig:limit} is somewhat different from the existing bound in the literature~\cite{Krnjaic:2015mbs}, although they agree at the order-of-magnitude  level. 			

The rest of the paper is organized as follows: in Section~\ref{sec:production} we derive the expression for energy loss due to the existence of $S$ and calculate the contributions from both the diagrams with $S$ coupling to nucleons and pions (see Fig.~\ref{fig:diagram}). All the relevant decays of $S$ are collected in Section~\ref{sec:decay}. The re-absorption rate of $S$ and the dependence of the corresponding MFP on $m_S$ and $\sin\theta$ are obtained in Section~\ref{sec:absorption}. The resultant supernova luminosity limit on $m_S$ and $\sin\theta$ and its complementarity with other existing and future laboratory limits are presented in Section~\ref{sec:limit}.  We conclude in Section~\ref{sec:conclusion}. Some of the calculational details of the emission rate have been relegated to Appendix~\ref{sec:appendix}.

\section{Production of $S$ in the supernova core}
\label{sec:production}
\begin{figure}[t!]
  \centering
  \includegraphics[width=0.4\textwidth]{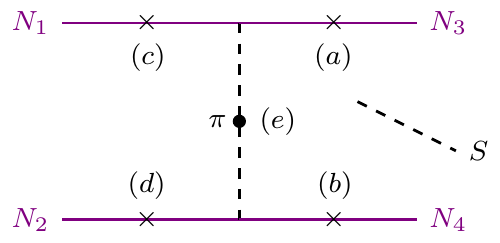}
  \includegraphics[width=0.4\textwidth]{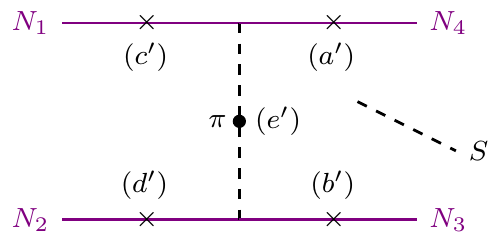}
  \caption{ Feynman diagrams for the production of light scalar $S$ in the nucleon bremsstrahlung process $N_1 + N_2 \to N_3 + N_4 +S$ in  the supernova core, with $N_i = p,\, n$. The left and right panels are respectively for the $t$ and $u$-channel. The light scalar $S$ can be attached to either of the nucleon lines $(a)$, $(b)$, $(c)$, $(d)$, $(a')$, $(b')$, $(c')$, $(d')$, as denoted by the crosses ($\times$), or to the pion mediator $(e)$, $(e')$, as denoted by the blobs ($\bullet$).}
  \label{fig:diagram}
\end{figure}

\begin{figure}[t!]
  \centering
  \includegraphics[width=0.7\textwidth]{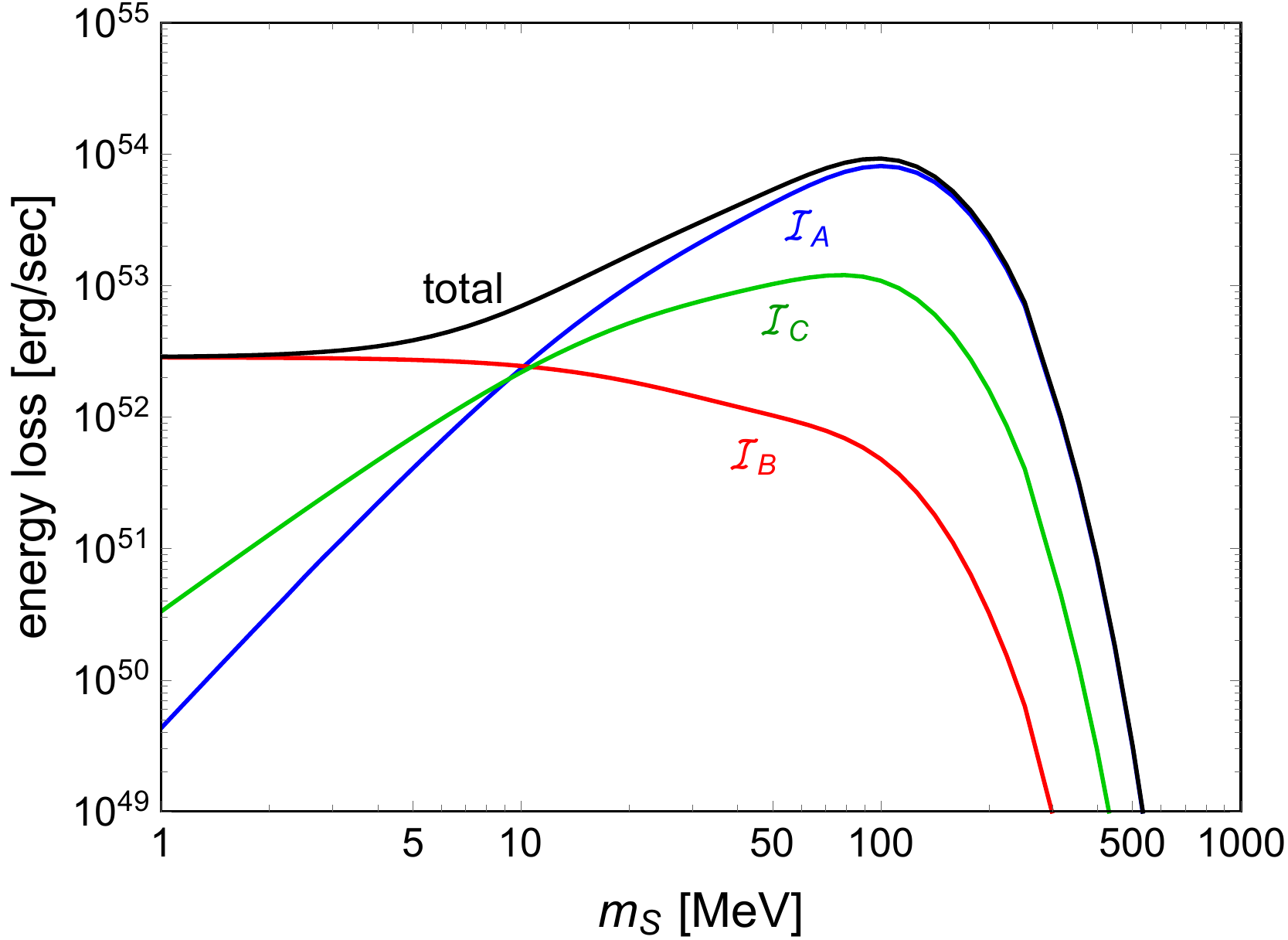}
  \caption{Contributions of the terms ${\cal I}_A$ (blue), ${\cal I}_B$ (red) and ${\cal I}_C$ (green) to energy loss in Eq.~(\ref{eqn:rate2}), respectively corresponding to the light scalar coupling to the nucleons, the pion mediators and the cross terms in Fig.~\ref{fig:diagram}. The black curve denotes the total contribution, as given by Eq.~\eqref{eq:Qtot}. Here we have set the supernova core temperature $T = 30$ MeV, nuclear density $n_B = 1.2 \times 10^{38}~{\rm cm}^{-3}$ and mixing angle $\sin\theta = 10^{-6}$ for illustration purposes.}
  \label{fig:comparison}
\end{figure}

As stated in the introduction, the dominant channel for the production of $S$ in the supernova core is through the nucleon bremsstrahlung process
\begin{eqnarray}
\label{eqn:2to3}
N + N \ \to \ N + N + S \,.
\end{eqnarray}
At the leading-order, this process is mediated by one-pion exchange (OPE) between nucleons, as shown in Fig.~\ref{fig:diagram}. In the calculation below we include both types of diagrams with $S$ coupling to nucleons~\cite{Shifman:1978zn, Cheng:1988im} and pions~\cite{Voloshin:1985tc, Donoghue:1990xh}. These diagrams are labeled by $(a)$, $(b)$, $(c)$, $(d)$, $(a')$, $(b')$, $(c')$, $(d')$ and $(e)$, $(e')$, and denoted respectively by the crosses and black blobs in Fig.~\ref{fig:diagram}.
As shown below, at the leading order in $m_S^2/m_N E_S$ (where $m_N$ is the nucleon mass and $E_S$ the energy of $S$), the Feynman diagrams $(a)$, $(b)$, $(c)$ and $(d)$, and $(a')$, $(b')$, $(c')$ and $(d')$ with $S$-nucleon couplings cancel out with each other, and we have to go to the next-to-leading order in $m_S^2/m_N E_S$ for these diagrams. As a result when the scalar mass $m_S \lesssim 10$ MeV, the contributions of $S$-nucleon coupling diagrams are suppressed by the scalar mass via $(m_S/E_S)^4$ [cf. Eq.~(\ref{eqn:Itot})], and the energy loss due to the existence of $S$ will be dominated by the $S$-pion coupling diagrams $(e')$ and $(e')$, as shown in Fig.~\ref{fig:comparison}.

%The scalar $S$ can couple to either of the nucleon lines or the pion mediators, which correspond respectively to the diagrams $(a)$, $(b)$, $(c)$, $(d)$, $(a')$, $(b')$, $(c')$, $(d')$ and $(e)$, $(e')$.
The $S$ couplings to both nucleons and pions are induced by  the mixing  of $S$ with the SM Higgs $h$, with the Lagrangian given by
\begin{eqnarray}
\label{eqn:lagrangian}
{\cal L} \ = \ \sin\theta S
\left[ y_{hNN} \overline{N}N + A_\pi (\pi^0\pi^0 + \pi^+ \pi^-) \right] \,,
\end{eqnarray}
where $y_{hNN}$ is the effective coupling of SM Higgs to nucleons. Through the couplings to quarks inside nucleons, it turns out that $y_{hNN} \simeq 10^{-3}$~\cite{Shifman:1978zn, Cheng:1988im}.
From chiral perturbation theory, the effective coupling of light $S$ to pions in Eq.~(\ref{eqn:lagrangian}) can be written as~\cite{Voloshin:1985tc, Donoghue:1990xh} %Monin:2018lee
\begin{eqnarray}
\label{eqn:Api}
{\cal A}_\pi \ = \ \frac{2}{9v_{\rm EW}}
\left( m_S^2+ \frac{11}{2} m_\pi^2 \right) \,,
\end{eqnarray}
with $v_{\rm EW} \simeq 246$ GeV is the electroweak vacuum expectation value. In order to discuss supernova limits, the relevant range of scalar mass is below the two-pion threshold, i.e. $m_S \lesssim 2m_\pi$, and it is clear in Eq.~(\ref{eqn:Api}) that the coupling ${\cal A}_\pi$ is highly suppressed by the ratio $m_\pi/v_{\rm EW} \sim 10^{-3}$ which is of the same order as $y_{hNN}$.

Following the calculations in Refs.~\cite{Giannotti:2005tn, Dent:2012mx, Kazanas:2014mca, Chang:2018rso, DeRocco:2019njg}, the energy emission rate per unit volume in the supernova core is given by
\begin{eqnarray}
\label{eqn:rate}
Q \ = \ \int {\rm d} \Pi_5 {\cal S} \sum_{\rm spins} |{\cal M}|^2 (2\pi)^4
\delta^4 (p_1 + p_2 - p_3 - p_4 - k_S) E_S f_1 f_2 P_{\rm decay} P_{\rm abs} \,,
\end{eqnarray}
where ${\rm d} \Pi_5$ is the $2\to 3$ phase space factor, ${\cal S}$ is the symmetry factor for identical particles, being $1$ for $np$ scattering process and $1/4$ for $pp$ and $nn$ processes, $\mathcal{M}$ is the scattering amplitude for $N + N \to N + N + S$, and $f_{1,\,2}$ the non-relativistic Maxwell-Boltzmann distributions of the two incoming nucleons in the non-degenerate limit, defined by
\begin{eqnarray}
f({\bf p}) \ = \ \frac{n_B}{2}\left(\frac{2\pi}{m_N T}\right)^{3/2}
{\rm{e}}^{-{\bf p}^2/2 m_N T} \,,
\end{eqnarray}
with $T \simeq 30$ MeV the supernova core temperature and $n_B \simeq 1.2 \times 10^{38}~{\rm cm}^{-3}$ the typical baryon number density in the supernova core. In Eq.~(\ref{eqn:rate}),
\begin{eqnarray}
\label{eqn:Rdecay}
P_{\rm decay} \ = \ {\rm e}^{-R_c \Gamma_S}
\end{eqnarray}
is the decay probability with $R_c$ being the supernova core size (only the $S$ particles decaying outside the supernova core will contribute to energy loss), and
\begin{eqnarray}
\label{eqn:Rabs}
P_{\rm abs} \ = \ {\rm e}^{- R_c /\lambda}
\end{eqnarray}
is the re-absorption probability, taking into account the fact that the light scalar $S$ can be re-absorbed inside the core via the $3\to 2$ inverse nucleon bremsstrahlung process
\begin{eqnarray}
\label{eqn:3to2}
N + N + S \ \to \ N + N \, ,
\end{eqnarray}
and $\lambda$ in Eq.~\eqref{eqn:Rabs} is the MFP of $S$.

The calculation details of the squared amplitude $\sum |{\cal M}|^2$ and the emission rate $Q$ are collected respectively in Appendices~\ref{sec:amplitude} and \ref{sec:rate}. It is interesting to notice that at the leading order in $m_N E_S \gg m_S^2$ [cf. Eq.~(\ref{eqn:LO})], the four $t$ and $u$-channel diagrams cancel out among themselves:
\begin{eqnarray}
\label{eqn:cancel}
\label{eqn:Mabcd}
{\cal M}_a + {\cal M}_b + {\cal M}_c + {\cal M}_d & \ \simeq \ & 0 \,, \\
\label{eqn:Mabcdp}
{\cal M}_{a'} + {\cal M}_{b'} + {\cal M}_{c'} + {\cal M}_{d'} & \ \simeq \ & 0 \,.
\end{eqnarray}
Incidentally, this cancellation  occurs only for the CP-even scalar case we are considering. For a vector boson (dark photon) or CP-odd scalar (axion/ALP) case, the Lorentz structures of the matrix elements are very different and no such cancellation occurs (see e.g. Refs.~\cite{Giannotti:2005tn, Dent:2012mx}). This is a key difference between the CP-even scalar and other kinds of light particles discussed in the literature that makes the CP-even case somewhat special.

Thus, to obtain a non-vanishing result for the sum of amplitudes $\sum |{\cal M}|^2$ in Eq.~\eqref{eqn:rate}, we have to go to next-to-leading order in the small parameter $m_S^2 / m_N E_S$.
%As a result, the production rate from the $S-N-N$ couplings is suppressed by $(m_S/E_S)^4$ (see Eq.~(\ref{eqn:Itot})).
After a long but straightforward calculation (cf.~Appendix~\ref{sec:appendix}), the emission rate in (\ref{eqn:rate}) can be simplified to
\begin{eqnarray}
\label{eqn:rate2}
Q & \ = \ & \frac{n_B^2 \alpha_\pi^2 f_{pp}^4 T^{7/2} \sin^2\theta}{8 \pi^{3/2} m_N^{9/2}}\int_q^\infty {\rm d}u \int_0^\infty {\rm d}v \int_{-1}^{1} {\rm d}z
\int_{q}^{\infty} {\rm d}x \ \delta (u-v-x) \nonumber \\
&&\qquad \qquad  \times
\sqrt{uv} e^{-u} x \sqrt{x^2 - q^2} P_{\rm decay} P_{\rm abs} {\cal I}_{\rm tot} \,,
\end{eqnarray}
where $\alpha_\pi \equiv (2m_N/m_\pi)^2/4\pi \simeq 15$, and the effective pion-nucleon coupling $f_{pp} \simeq 1$, and $u$, $v$, $z$, $x$, $q$ are dimensionless parameters defined in Eq.~(\ref{eqn:uvxyqz}).  The dimensionless function
\begin{eqnarray}
\label{eqn:Itot}
{\cal I}_{\rm tot} \ = \
y_{hNN}^2 \left( \frac{q}{x} \right)^4 {\cal I}_A +
\frac{1}{81} \left( \frac{m_N}{v_{\rm EW}} \right)^2 {\cal I}_B +
\frac{2}{9} y_{hNN} \left( \frac{q}{x} \right)^2
\left( \frac{m_N}{v_{\rm EW}} \right) {\cal I}_C \,,
\end{eqnarray}
where we have summed up the contributions from the $pp$, $nn$ and $np$ processes:
\begin{eqnarray}
\label{eqn:Iabc}
{\cal I}_{A,\, B,\, C} \ = \ {\cal I}_{A,\, B,\, C}^{(pp)} + {\cal I}_{A,\, B,\, C}^{(nn)} + 4{\cal I}_{A,\, B,\, C}^{(np)} \,,
\end{eqnarray}
with ${\cal I}_{A,\, B,\, C}^{(pp)} = {\cal I}_{A,\, B,\, C}^{(nn)}$.
For the $pp$ and $nn$ channels, there is a factor of $1/2 \times 1/2  = 1/4$ for identical particles in both the initial and final states, thus we have an extra factor of $4$ for the $np$ term in Eq.~(\ref{eqn:Iabc}). The dimensionless functions ${\cal I}_{A,\, B,\, C}^{(pp,\,np)}$ can be found in Eqs.~(\ref{eqn:IApp}) to (\ref{eqn:ICnp}) in Appendix~\ref{sec:rate}. In Eq.~(\ref{eqn:Itot})
%the functions ${\cal I}_{A,\,B,\,C}$ can be found in Eq.~(\ref{eqn:Iabc}) in Appendix~???.
the ${\cal I}_A^{}$ term denotes the contribution from the diagrams $(a)$, $(b)$, $(c)$, $(d)$ and $(a')$, $(b')$, $(c')$, $(d')$, with the scalar $S$ coupling to the nucleons. Besides the small Yukawa coupling $y_{hNN}$, the function ${\cal I}_A$ is also suppressed by the scalar mass $m_S$ via $(q/x)^4 = (m_S/E_S)^4$ when $S$ is light (note that the scalar energy $E_S \sim 3T$ when $m_S \ll 3T$).
%\blue{Should not the scalar eenergy be 3T in the relativistic casee?}
The ${\cal I}_B^{}$ function is the contribution from the diagrams $(e)$ and $(e')$ with $S$ coupling to the pion mediator, and is highly suppressed by the ratio $(m_N/9v_{\rm EW})^2$. The ${\cal I}_C^{}$ term stands for the interference contributions from the two sets of diagrams above, and is expected to be always in between ${\cal I}_A$ and ${\cal I}_B$. This is explicitly shown in Fig.~\ref{fig:comparison} with a benchmark value of the mixing angle $\sin\theta=10^{-6}$.
%In Eq.~(\ref{eqn:Itot}) the functions ${\cal I}_{A,\,B,\,C}$ .

%which correspond respectively to the ${\cal I}_{A,\, B,\, C}$ terms in Eq.~(\ref{eqn:Itot}).
In Fig.~\ref{fig:comparison}, the ${\cal I}_{A,\,B,\,C}$ contributions to energy loss as functions of the scalar mass $m_S$ are presented respectively by the blue, red and green curves, while the black curve is the total contribution
\begin{eqnarray}
{\cal Q} \ = \ Q V_c \,,
\label{eq:Qtot}
\end{eqnarray}
where $V_c = 4\pi R_c^3/3$ is the supernova core volume with the core radius $R_c \simeq 10$ km. As expected in Eq.~(\ref{eqn:Itot}) and as can be seen from Fig.~\ref{fig:comparison}, the ${\cal I}_A$ term is suppressed by the scalar mass for $m_S \lesssim 100$ MeV, via $(q/x)^4 = (m_S/E_S)^4$. It originates from the cancellation in Eq.~(\ref{eqn:cancel}). The ${\cal I}_B$ contribution is almost a constant in the limit of small scalar mass $m_S$, and dominates over other terms for $m_S \lesssim 10$ MeV, although it is suppressed by $(m_N/v_{\rm EW})^2$, as can be seen from Eq.~(\ref{eqn:Itot}). As expected, the contribution of the cross term ${\cal I}_C$ is always in between ${\cal I}_A$ and ${\cal I}_B$. Also all the three contributions are Boltzmann-suppressed in the limit of large scalar mass $m_S\gtrsim 300$ MeV.

\section{Decay of $S$}
\label{sec:decay}

From mixing with the SM Higgs, the light scalar $S$ can decay into the SM particles. The supernova limits are only relevant if the scalar mass is $\lesssim 300$ MeV; otherwise the production rate will be highly Boltzmann-suppressed (see Fig.~\ref{fig:comparison}). Therefore only the following decay channels are relevant:
\begin{eqnarray}
\Gamma_0 (S \to e^+e^-) & \ = \ &
\frac{m_S m_e^2 \sin^2\theta}{8\pi v_{\rm EW}^2}
\left( 1-\frac{4m_e^2}{m^2_{S} }\right)^{3/2} \,, \label{eqn:See}\\
\Gamma_0 (S \to \mu^+\mu^-) & \ = \ &
\frac{m_S m_\mu^2 \sin^2\theta}{8\pi v_{\rm EW}^2}
\left( 1-\frac{4m_\mu^2}{m^2_{S} }\right)^{3/2} \,, \\
\label{eqn:Saa}
\Gamma_0 (S \to \gamma\gamma) & \ = \ &
\frac{\alpha^2 m_{S}^3 \sin^2\theta}{256\pi^3 v_{\rm EW}^2}
\left| \sum_f N_C^f Q_f^2 A_{1/2} (\tau_f) + A_1 (\tau_W) \right|^2 \,, \\
\Gamma_0 (S \to \pi^0 \pi^0) & \ = \ &
\frac{\sin^2\theta}{648\pi m_S v_{\rm EW}^2}
\left( m_S^2+ \frac{11}{2} m_{\pi^0}^2 \right)^2
\left( 1-\frac{4m_{\pi^0}^2}{m_S^2} \right)^{1/2} \,,  \\
\Gamma_0 (S \to \pi^+ \pi^-) & \ = \ &
\frac{\sin^2\theta}{324\pi m_S v_{\rm EW}^2}
\left( m_S^2+ \frac{11}{2} m_{\pi^\pm}^2 \right)^2
\left( 1-\frac{4m_{\pi^\pm}^2}{m_S^2} \right)^{1/2} \,,
\label{eqn:Spipi}
\end{eqnarray}
with $m_{e,\,\mu,\, \pi^0,\, \pi^\pm}$ being respectively the masses of $e$, $\mu$, $\pi^0$ and $\pi^\pm$. In Eq.~(\ref{eqn:Saa}) we include only the SM fermions and $W$ bosons running in the loop for the diphoton channel, with $\alpha=e^2/4\pi$ the fine-structure constant, $N_C=3~(1)$ the color factor for quarks (charged leptons), $\tau_X=m_S/4m_X^2$, and the loop functions $A_{1/2}(\tau_f),~A_1(\tau_W)$ can be found in Appendix A of Ref.~\cite{Dev:2017dui}.  Note that all the couplings are universally from mixing with the SM Higgs, i.e. proportional to the mixing angle $\sin\theta$. As mentioned in the introduction, if there are other beyond SM heavy particles in the loop which couple both to the scalar $S$ and photons, this might dramatically change the decay (and production) of $S$ in the supernova core (see e.g. Ref.~\cite{Dev:2019hho}).
%The left-right symmetric model is such an example~\cite{Pati:1974yy, Mohapatra:1974gc, Senjanovic:1975rk}, where the light scalar $S$ originates form the $SU(2)_R$-breaking triplet, and the heavy $W_R$ boson and heavy scalars could mediate the decay $S \to \gamma\gamma$ even if the mixing angle $\sin\theta$ vanishes~\cite{Nemevsek:2012cd, Dev:2016vle, Dev:2017dui}.
For simplicity, we assume that no such heavy particles exist throughout this paper.

In addition to the two-body decays above, if kinematically allowed, the light scalar $S$ has also the following four-body decays via the SM $W$ and $Z$ bosons:
\begin{eqnarray}
&& S \ \to\  W^{+ \, \ast} W^{- \, \ast} \ \to \ \ell_\alpha^+ \ell_\beta^{-} \nu_\alpha \bar\nu_\beta \quad \text{with} \quad
\alpha,\,\beta = e,\, \mu \,,  \\
&& S \ \to \ Z^{\ast} Z^{\ast} \ \to \ \nu \nu \bar\nu \bar\nu \,, \;\;\;
e^+ e^- e^+ e^- \,, \;\;\; e^+ e^- \mu^+ \mu^- \,.
\end{eqnarray}
The four-body decay widths can be calculated, e.g. following the procedure in the Appendix of Ref.~\cite{Dev:2018kpa}. Suppressed by the phase space, it turns out that the partial widths of the four-body decays are much smaller than the two-body decays above, and can be safely neglected. Then we can insert the following width into Eq.~(\ref{eqn:Rdecay}):
\begin{eqnarray}
%R_{\rm decay} = \exp\{ - R_c \Gamma_S \} \,, \quad \text{with} \quad
\Gamma_S & \ = \ & \frac{m_S}{E_S} \Gamma_{S,\,0} \,,
\end{eqnarray}
where $\Gamma_{S,\,0}$ is the proper total width, i.e. the sum of all partial widths listed in Eqs.~\eqref{eqn:See}-\eqref{eqn:Spipi}, and $m_S/E_S$ the inverse Lorentz boost factor.

\begin{figure}[!t]
  \centering
  \includegraphics[width=0.65\textwidth]{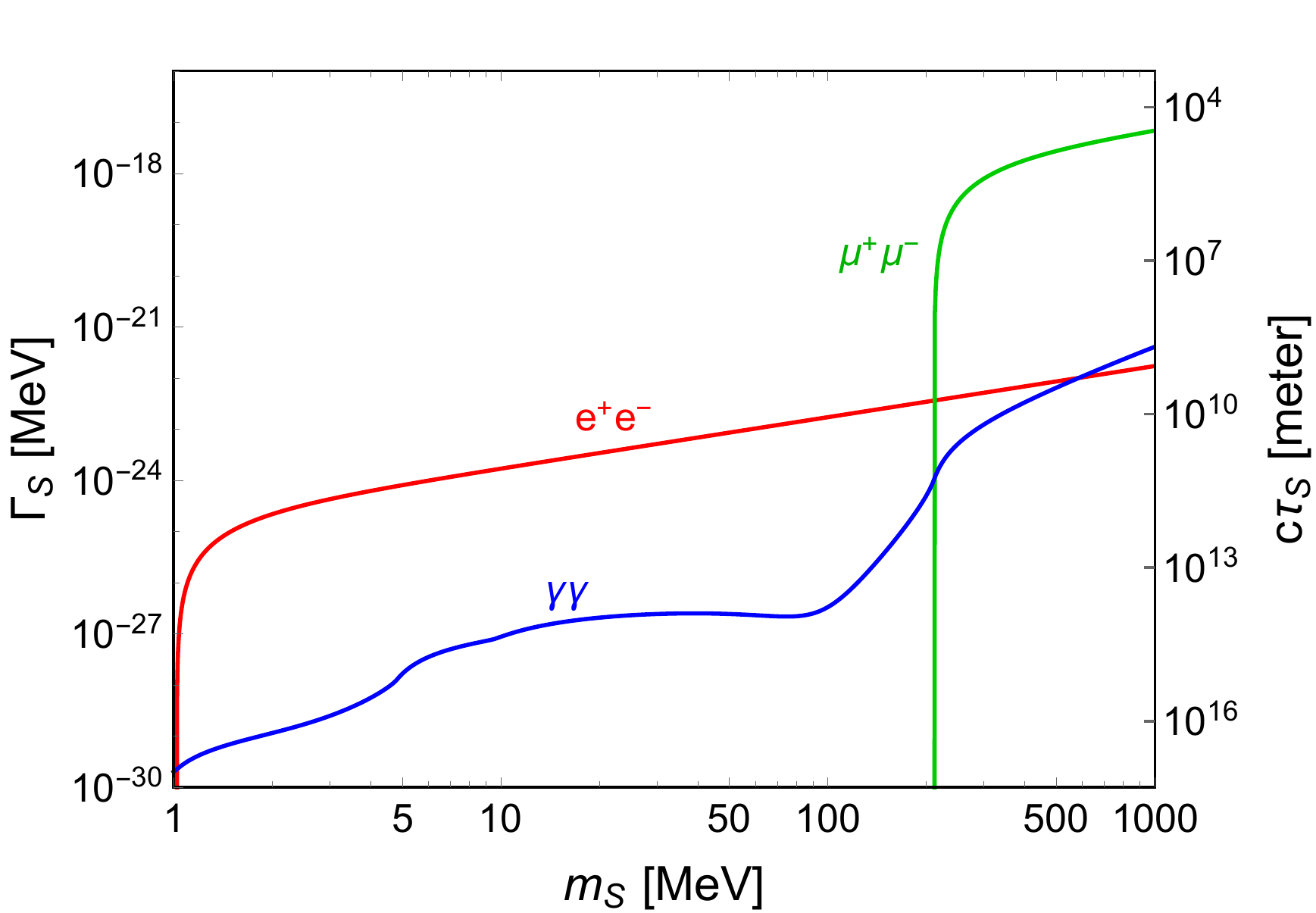}
  \caption{Decay widths of $S$ into $e^+ e^-$ (red), $\mu^+ \mu^-$ (green) and $\gamma\gamma$ (blue), with $\sin\theta = 10^{-6}$. We also show the corresponding proper decay length $c\tau_S$.}
  \label{fig:width}
\end{figure}

The partial widths of $S$ into $e^+ e^-$, $\mu^+ \mu^-$ and $\gamma\gamma$ and the corresponding proper decay length $c\tau_S$ (where $\tau_S=1/\Gamma_S$) are shown in Fig.~\ref{fig:width} respectively by the red, green and blue curves. For concreteness we have set the mixing angle to be the benchmark value of $\sin\theta = 10^{-6}$. It is clear from Fig.~\ref{fig:width} that the decay of $S$ is dominated by $e^+ e^-$ channel when the scalar mass $m_S \lesssim 2 m_\mu \simeq 210$ MeV and by the decay $\mu^+ \mu^-$ above the $2m_\mu$ threshold. When the scalar $S$ is heavier than the two-pion threshold $2m_\pi \simeq 270$ MeV, it will decay mostly into pions. However, above the $2m_\pi$ threshold the scalar $S$ decays so fast that the range $m_S \gtrsim 270$ MeV can no longer be excluded by supernova energy loss criteria (see Fig.~\ref{fig:limit}). Therefore we do not show the partial decay width of $S$ into pions in Fig.~\ref{fig:width}.

\section{Re-absorption of $S$}
\label{sec:absorption}

As mentioned earlier, after being produced, the light scalar $S$ may be re-absorbed via the inverse nucleon bremsstrahlung process in Eq.~(\ref{eqn:3to2}). As a result, the scalar $S$ can only travel a finite distance in the supernova core, and we can define the corresponding inverse MFP as~\cite{Burrows:1990pk, Giannotti:2005tn}
\begin{eqnarray}
\label{eqn:mfp}
\lambda^{-1} & \ \equiv \ &
\frac{1}{2 E_S} \frac{d {\cal N}_S (-k_S)}{d \Pi_S} \nonumber \\
& \ = \ &
\frac{1}{2 E_S}
\int {\rm d} \Pi_4 {\cal S} \sum_{\rm spins} |{\cal M}'|^2 (2\pi)^4
\delta^4 (p_1 + p_2 - p_3 - p_4 + k_S) f_1 f_2 \,,
\end{eqnarray}
with ${\cal N}_S (-k_S)$ the total $S$ number per unit volume produced in the supernova core with momentum $-k_S$ (note that for the re-absorption process the scalar $S$ is in the initial state, while in the production process $S$ is in the final state), $\Pi_S$ the phase space of $S$, ${\rm d}\Pi_4$ the four-body phase space for the initial and final state nucleons, and ${\cal M}^\prime$ the scattering amplitude for the inverse  nucleon bremsstrahlung process in Eq.~(\ref{eqn:3to2}). It is straightforward to prove that
\begin{eqnarray}
\sum_{\rm spins}\left| {\cal M} (k_S) \right|^2 \ = \
\sum_{\rm spins}\left| {\cal M}^\prime (-k_S) \right|^2 \,.
\end{eqnarray}
Following the same procedure as above for the emission rate $Q$, the twelve-dimensional integration can be greatly simplified:
\begin{eqnarray}
\label{eqn:mfp2}
\lambda^{-1} & \ = \ &
\frac{n_B^2}{64 \pi^{3/2} m_N^{5/2} T^{1/2} x}
\int_0^\infty {\rm d}u \int_q^\infty {\rm d}v \int_{-1}^{1} {\rm d}z \sqrt{uv} e^{-u}
\delta (u-v+x) {\cal S} \sum_{\rm spins} |{\cal M}|^2 \nonumber \\
& \ = \ & \frac{n_B^2 \pi^{1/2} \alpha_\pi^2 f_{pp}^4 \sin^2\theta}{4m_N^{9/2} T^{1/2} x}
\int_0^\infty {\rm d}u \int_q^\infty {\rm d}v \int_{-1}^{1} {\rm d}z \sqrt{uv} e^{-u}
\delta (u-v+x) {\cal I}_{\rm tot} \,,
\end{eqnarray}
with the dimensionless variables $u$, $v$, $z$, $x$ and the dimensionless function ${\cal I}_{\rm tot}$ being the same as in Eq.~(\ref{eqn:rate2}).\footnote{Note that in Eqs.~(\ref{eqn:rate2}) and (\ref{eqn:mfp2}) the integration  ranges for the dimensionless parameters $u$ and $v$ are different. For the bremsstrahlung production process, the scalar $S$ is in the final state, so we require $u>q$ such that the $S$ production is kinematically allowed. On the other hand, for the absorption process, the scalar $S$ is in the initial state, and as a result, $v> q$ but $u$ can be smaller. This is also ensured by the arguments in the delta-functions in the two equations. } It is obvious from Eq.~(\ref{eqn:mfp2}) that the MFP $\lambda$ not only depends on the scalar mass $m_S$ and mixing angle $\sin\theta$, but also on the scalar energy $E_S = x T$ (note that in Eq.~(\ref{eqn:mfp2}) the variable $x$ is not integrated out, unlike in Eq.~(\ref{eqn:rate2})). To simplify the calculation of emission rate $Q$ in Eq.~(\ref{eqn:rate2}), we average over the distributions of $S$ and replace the MFP $\lambda$ in the $P_{\rm abs}$ factor in Eq.~(\ref{eqn:Rabs}) by the effective energy-independent MFP $\langle \lambda \rangle$, with its inverse defined by~\cite{Ishizuka:1989ts}
\begin{eqnarray}
\langle \lambda^{-1} \rangle \ \equiv \
\frac{\bigintss {\rm d} E_S \frac{E_S^3}{e^{E_S/T}-1} \lambda^{-1} (E_S)}{\bigintss {\rm d} E_S \frac{E_S^3}{e^{E_S/T}-1}}
\ = \ \frac{\bigintss {\rm d} x \frac{x^3}{e^{x}-1} \lambda^{-1} (x)}{\bigintss {\rm d} x \frac{x^3}{e^{x}-1}} \,.
\end{eqnarray}

\begin{figure}[!t]
  \centering
  \includegraphics[width=0.7\textwidth]{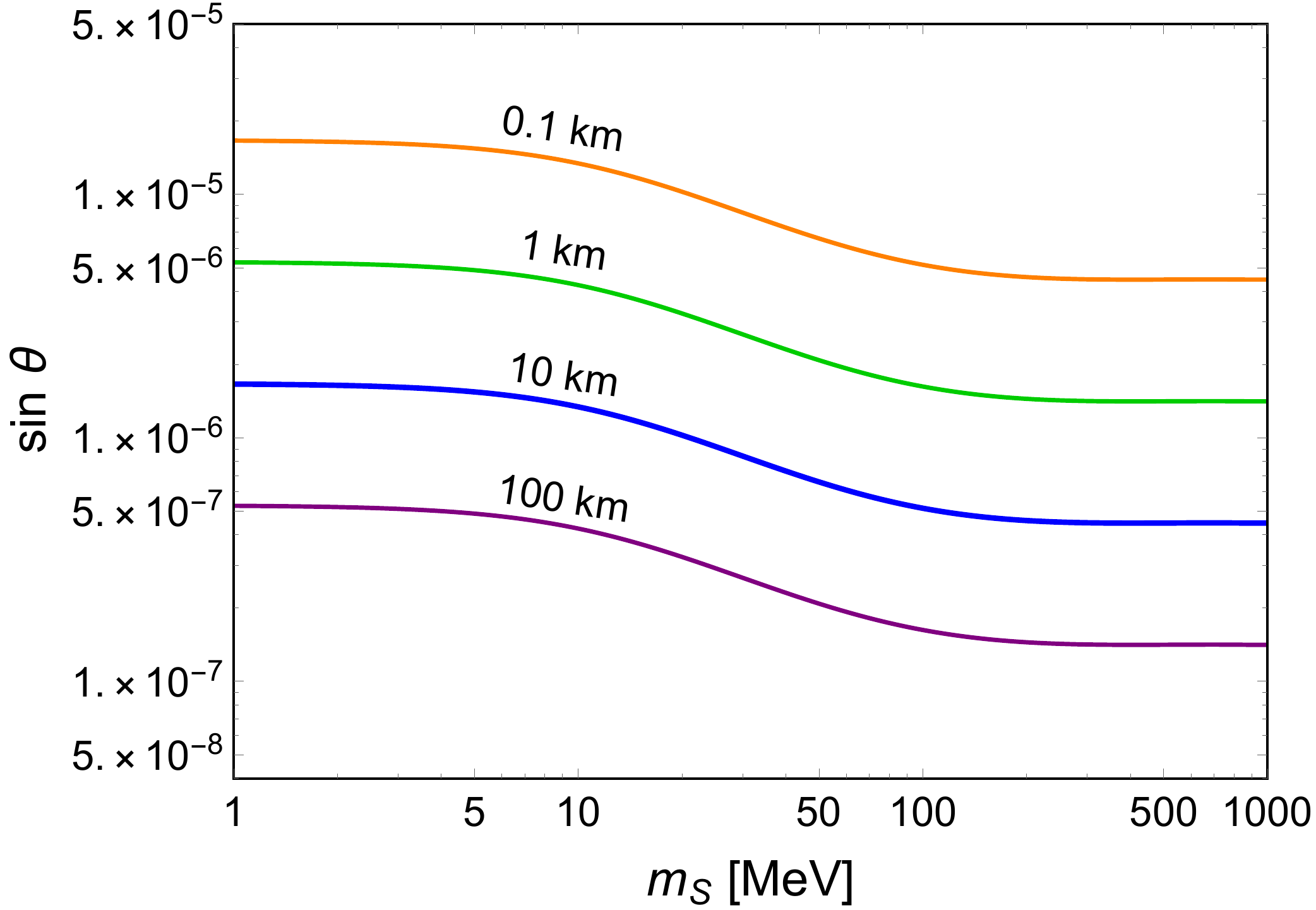}
  \caption{Dependence of effective MFP $\langle \lambda \rangle$ on the scalar mass $m_S$ and mixing angle $\sin\theta$. The orange, green, blue and purple lines correspond respectively to $\langle \lambda \rangle = 0.1$ km, 1 km, 10 km and 100 km. We take the supernova core temperature $T = 30$ MeV and nuclear density $n_B = 1.2 \times 10^{38}~{\rm cm}^{-3}$.}
  \label{fig:mfp}
\end{figure}

The dependence of the effective MFP on the scalar mass $m_S$ and mixing angle $\sin\theta$ is shown in Fig.~\ref{fig:mfp}, where the orange, green, blue and purple lines correspond respectively to the mean free path values of 0.1 km, 1 km, 10 km and 100 km.  From Fig.~\ref{fig:mfp}, one can make a good estimate of the order-of-magnitude MFP by the following empirical formula:
\begin{eqnarray}
\langle\lambda\rangle \ \simeq \ 10 \; {\rm km} \times \left( \frac{\sin\theta}{10^{-6}} \right)^{-2} \,.
\end{eqnarray}
This can be verified from the prefactor in Eq.~({\ref{eqn:mfp2}}) and the fact that ${\cal I}_{\rm tot} \sim 10^{-6}$ [cf. Eq.~(\ref{eqn:Itot})].
%As expected in Eq.~(\ref{eqn:mfp2}), the MFP $\lambda \propto (\sin^\theta)^{-2}$.
%\blue{\sout{In the limit of large scalar mass, $\lambda^{-1}$ is highly suppressed by the $f_1 f_2$ factor in Eq.~(\ref{eqn:mfp}); therefore in Fig.~\ref{fig:mfp} the MFP $\langle\lambda\rangle$ is largely enhanced for $m_S \gtrsim 300$ MeV.}}

\section{Luminosity limit}
\label{sec:limit}

%(i) the light scalar couples to the nucleons, with the diagrams $(a)$, $(b)$, $(c)$, $(d)$ and $(a')$, $(b')$, $(c')$, $(d')$ in Fig.~\ref{fig:diagram}, which corresponds to the term ${\cal I}_A$ in Eq.~(\ref{eqn:rate2}), (ii) the scalar $S$ couples to the pion, which corresponds to the diagrams $(e)$ and $(e')$ and corresponds to the ${\cal I}_B$ term, and (iii) the cross terms between the two sets of diagrams above, i.e. the ${\cal I}_C$ term.

\begin{figure}[!t]
  \centering
  \includegraphics[width=0.7\textwidth]{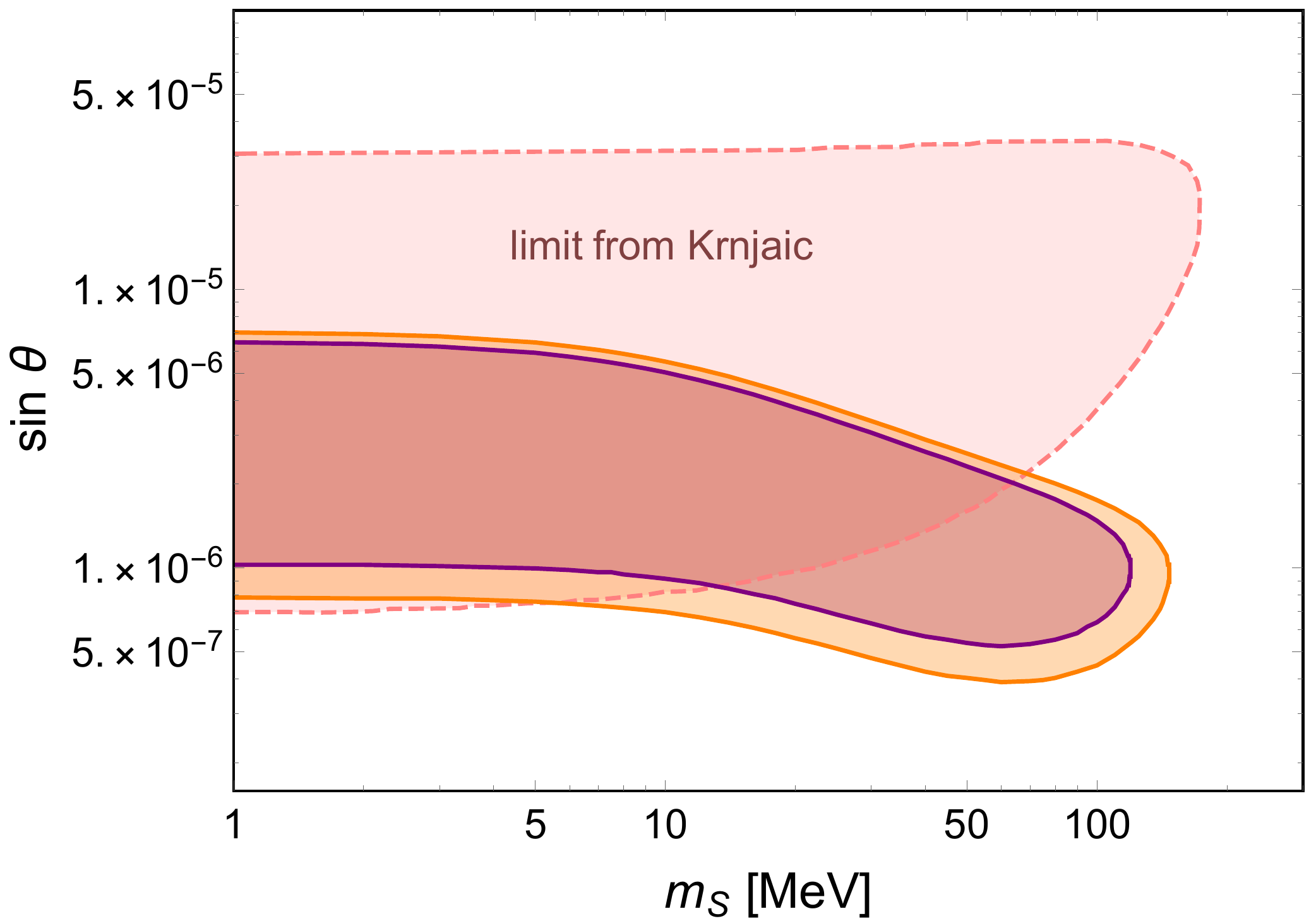}
  \caption{Supernova limit on the scalar mass $m_S$ and mixing angle $\sin\theta$. The orange and purple shaded regions respectively correspond  to the luminosity limit of $3\times10^{53}$ erg/sec and $5\times10^{53}$ erg/sec. For comparison, the pink region surrounded by dashed line is the limit from Ref.~\cite{Krnjaic:2015mbs}. We have set supernova core temperature $T = 30$ MeV and nuclear density $n_B = 1.2 \times 10^{38}~{\rm cm}^{-3}$.}
  \label{fig:limit}
\end{figure}

With the decay and re-absorption effects in Section~\ref{sec:decay} and \ref{sec:absorption}, we are now ready to calculate the emission rate $Q$ in Eq.~(\ref{eqn:rate2}) in the supernova core due to the existence of the light scalar $S$.  Based on the observed neutrinos from SN1987A~\cite{Hirata:1987hu}, the energy loss ${\cal L}_\nu$ due  to neutrino emission is expected to be $(3 - 5) \times 10^{53}$ erg/sec~\cite{Raffelt:1996wa} and can be used to set limits on the scalar mass $m_S$ and the mixing angle $\sin\theta$. Setting the total energy emission rate ${\cal Q}$ to be $3\times 10^{53}$ erg/sec and $5\times 10^{53}$ erg/sec, we can exclude the orange and purple shaded regions in the $m_S$ and $\sin\theta$ parameter space  in Fig.~\ref{fig:limit} respectively. Here we have taken a fixed supernova core temperature of $T=30$ MeV and a constant nuclear core density of $n_B=1.2\times 10^{38}~{\rm cm}^{-3}$ for simplicity, although there might be some uncertainty in these parameters due to the uncertainty in the progenitor proto-neutron star mass and neutrino flux estimations~\cite{Keil:2002in, Chang:2016ntp, Chang:2018rso}.

{For comparison, the previous limit from Ref.~\cite{Krnjaic:2015mbs} with the same supernova parameters is shown as the pink shaded region. Our result agrees with Ref.~\cite{Krnjaic:2015mbs} at an order-of-magnitude level, but disagrees in details. We believe this difference can be mainly attributed to the fact that the energy loss rate in Ref.~\cite{Krnjaic:2015mbs} [cf. his Eq.~(34)] was directly taken from Ref.~\cite{Ishizuka:1989ts} [cf. their Eq.~(20)], appended by an additional $\xi$-factor to approximately account for the finite scalar mass effect. However, we notice two issues with the analysis of Ref.~\cite{Ishizuka:1989ts} for the CP-even scalar (their `dilaton' $D$) case:
\begin{enumerate}
\item [(i)] They missed the cancelation effect in the nucleon contributions coming from different diagrams in Fig.~\ref{fig:diagram} (see our Appendix~\ref{sec:amplitude} for details), which is special to the CP-even scalar case.  This gives rise to an additional $(m_S/E_S)^4$ factor in the rate of production of $S$ via its coupling to nucleons.
\item [(ii)] They did not include the contribution from the diagrams with scalar coupling to the pion mediator (see Fig.~1 in Ref.~\cite{Ishizuka:1989ts} versus our Fig.~\ref{fig:diagram}). As shown in our Fig.~\ref{fig:comparison}, the pion contribution $\mathcal{I}_B$, together with the cross term $\mathcal{I}_C$, dominates over the nucleon contribution $\mathcal{I}_A$ for relatively smaller scalar masses.
\end{enumerate}
Therefore, the updated limit presented here should supersede the old limit from Ref.~\cite{Krnjaic:2015mbs}.}

As shown in Fig.~\ref{fig:limit}, there is an upper ``cut-off'' of the scalar mass $m_S$ around 147 MeV, as when the scalar mass is larger than the temperature the production rate is exponentially suppressed and the scalar also tends to decay faster such that it can not effectively take away energy from the supernova core. The excluded mixing angles range from $3.9\times10^{-7}$ to $7.0\times10^{-6}$ for the luminosity limit of $3\times 10^{53}$ erg/sec and  from $5.2\times10^{-7}$ to $6.4\times10^{-6}$ for $5\times 10^{53}$ erg/sec, depending on the scalar mass. The shape of excluded regions in Fig.~\ref{fig:limit} is a combined effect of the superposition of the three contributions ${\cal I}_{A,\, B,\, C}$, the decay of $S$, and the dependence of effective MFP $\langle\lambda\rangle$ on the scalar mass and mixing angle. One can expect that when large-scale neutrino detectors like IceCube-DeepCore~\cite{Collaboration:2011ym, Aartsen:2013nla}, Hyper-Kamiokande~\cite{Seo:2017cyz, Abe:2018uyc} and DUNE~\cite{Ankowski:2016lab, Abi:2020evt}  collect more neutrinos from the explosion of a nearby supernovae, the luminosity limits on the scalar mass $m_S$ and mixing angle can be significantly improved.

%The dashed gray line Fig.~\ref{fig:limit} corresponds to the limit without the re-absorption factor $R_{\rm abs}$ in Eq.~(\ref{eqn:rate2}), with limit set to be $5\times 10^{53}$ erg/sec.

\begin{figure}[!t]
  \centering
  \includegraphics[width=0.7\textwidth]{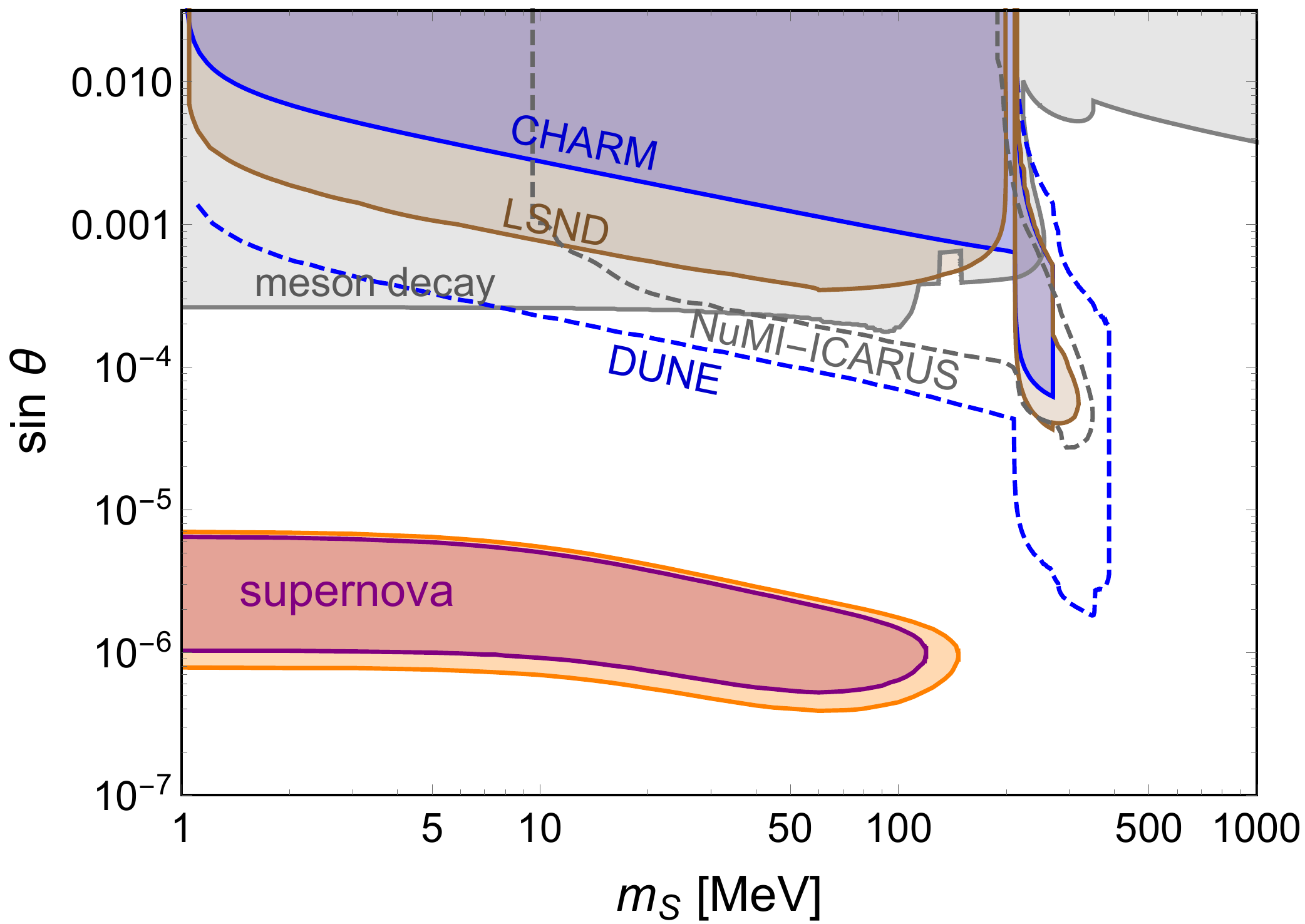}
  \caption{Complementarity of the supernova limits on $m_S$ and $\sin\theta$ (shaded purple and orange, cf. Fig.~\ref{fig:limit}) and those from FCNC meson decays (shaded gray), LSND (shaded brown), CHARM (shaded blue). The dashed gray and blue curves show respectively the future prospects at NuMI-ICARUS and DUNE.
  %\BD{Include ICARUS limit from \url{https://arxiv.org/pdf/1909.11670.pdf} Fig. 15, right panel, NuMI-ICARUS.}
  }
  \label{fig:complementary}
\end{figure}

%\section{Comparison with Existing Constraints} \label{sec:comp}
The supernova luminosity limits on the scalar mass $m_S$ and mixing angle $\sin\theta$ obtained here are largely complementary to other laboratory, astrophysical and cosmological limits, as summarized in Fig.~\ref{fig:complementary}. Through mixing with the SM Higgs, the scalar $S$ could obtain flavor-changing neutral current (FCNC) couplings to the SM quarks at 1-loop level, such as $S \bar{s}d$. These couplings induce the FCNC decays $K \to \pi +X$, $B \to K + X$ and $B \to \pi + X$ with $X = e^+ e^-,\, \mu^+ \mu^-,\, \gamma\gamma$ or missing energy. As the corresponding branching fractions are all highly suppressed in the SM, these decays provide the most stringent laboratory limits on the light scalar $S$. Combining the limits from NA48/2~\cite{Batley:2009aa,Batley:2011zz}, E949~\cite{Artamonov:2009sz}, KOTO~\cite{Ahn:2018mvc}, NA62~\cite{Ceccucci:2014oza,Ruggiero:2019}, KTeV~\cite{AlaviHarati:2003mr,AlaviHarati:2000hs,Alexopoulos:2004sx, Abouzaid:2008xm}, BaBar~\cite{Aubert:2003cm,Lees:2013kla}, Belle~\cite{Wei:2009zv}, and LHCb~\cite{Aaij:2012vr}, the meson decay limits are shown as the gray shaded regions in Fig.~\ref{fig:complementary}. The calculation details can be found e.g. in Refs.~\cite{Dev:2017dui,Dev:2019hho}. For the scalar $S$ with mass $40 \; {\rm MeV} \lesssim m_S \lesssim 350$ MeV, the meson decay limits can be improved to some extent by NuMI-ICARUS~\cite{Batell:2019nwo}, as indicated by the dashed gray line in Fig.~\ref{fig:complementary}.

The scalar $S$ can also be produced by proton bremsstrahlung in the LSND experiment, and the current LSND electron and muon data~\cite{Aguilar:2001ty, Auerbach:2003fz} have excluded the brown shaded regions in Fig.~\ref{fig:complementary}~\cite{Foroughi-Abari:2020gju}. The scalar $S$ might also be produced in the high-intensity beam-dump experiments. The current most stringent limits are from kaon decays in the CHARM experiment~\cite{Bergsma:1985qz}, which is presented as the blue shaded region in Fig.~\ref{fig:complementary} (see e.g. Refs.~\cite{Dev:2017dui,Dev:2019hho} for more details). With a higher luminosity, the DUNE near-detector experiment~\cite{Acciarri:2016ooe} can improve the CHARM limit by over an order of magnitude~\cite{Berryman:2019dme}, as indicated by the dashed blue line (corresponding to $\geq 10$ events).

Let us now make a few brief comments on the big bang nucleosynthesis (BBN) constraints on a light CP-even scalar. Under the two conditions of (i) $S$ is in thermal equilibrium with the SM particles at the BBN temperature of MeV scale~\cite{Dev:2017dui}, and (ii) the lifetime $\tau_S \gtrsim 1$ sec~\cite{Kainulainen:2015sva, Fradette:2017sdd}, it will  contribute an extra degree of freedom and spoil the success of BBN. The equilibrium condition can be estimated by comparing the $S$ interaction rates with the Hubble rate at BBN temperature.  It turns out that there is almost no parameter space of $m_S$ that satisfies both conditions (i) and (ii) simultaneously.  As a result the BBN limit on $S$ is expected to be very weak and thus not shown in Fig.~\ref{fig:complementary}. Alternatively, to maintain the success of BBN, the $S$, if already in equilibrium, must decouple above the QCD phase transition. More detailed analysis of the BBN limit is beyond the scope of this paper, and will be pursued in a future publication. Similarly, a sufficiently light scalar $S$ in thermal equilibrium could also contribute to the light degrees of freedom $N_{\rm eff}$ in the early universe and thus be constrained by the precision Planck data~\cite{Aghanim:2018eyx}. However, the limit from $\Delta N_{\rm eff}$ is very weak, i.e.~$\sin\theta \lesssim 0.01$~\cite{Dev:2019hho}, hence it is not shown in Fig.~\ref{fig:complementary}.

%especially for very small mixing angles, which might render $S$ decoupled much earlier.

%For sufficiently small mixing angle $\sin\theta$, \blue{as long as the scalar $S$ is in thermal equilibrium in the early universe~\cite{Dev:2017dui} and its lifetime $\tau_S \gtrsim 1$ sec}, then it will  \blue{

%The corresponding limit on $m_S$ and $\sin\theta$ is shown as the pink shaded region in Fig.~\ref{fig:complementary}. More details can be found e.g. in Ref.~\cite{Fradette:2017sdd}, but it requires further analysis beyond the scope of this paper to derive the full validity region of the BBN limit, especially for very small mixing angles, which might render $S$ decoupled much earlier.

Taking into consideration all the existing laboratory, astrophysical and cosmological limits on the light CP-even scalar $S$ and its mixing angle $\sin\theta$ with the SM Higgs, it is clear in Fig.~\ref{fig:complementary} that the luminosity limits from SN1987A are one or two orders of magnitude lower than the current laboratory constraints, depending on the scalar mass.
%It turns out that the supernova constraint does not exclude any additional parameter space not yet excluded by the BBN constraint for the range of scalar masses shown in this Figure. However, as mentioned above, the BBN constraint will not be applicable for much smaller $m_S$ (with the $\sin\theta$ values relevant for the supernova constraint), as the $S$ tends to be decoupled earlier than the QCD epoch, whereas the supernova limit shown here is valid all the way down to $m_S\to 0$.
%\blue{\sout{It is remarkable that the future DUNE experiment could reach the supernova excluded regions for $m_S \gtrsim 200$ MeV.}}

\section{Conclusion}
\label{sec:conclusion}

Due to their high nucleon density, supernova cores provide an ideal testing ground for light hypothetical particles, such as axion/ALP, dark photon or a light CP-even scalar, weakly coupling to the SM particles. Taking into consideration the core energy loss from the emission of these particles, stringent limits on their couplings to the SM can be derived. While there has been considerable literature on this for the dark photon and axion cases, the discussion of  supernova limits on a light CP-even scalar is very limited. Motivated to fill this gap, we have presented a complete calculation of the supernova luminosity limit on a generic light CP-even scalar with mass $m_S$ and mixing angle $\sin\theta$ with the SM Higgs. We consider the scalar production via the nucleon bremsstrahlung process $N N \to N N S$ in the supernova core and point out that, as a result of the distinct Lorentz structure of the couplings, the supernova limits on the CP-even scalar $S$ are very different from those on the dark photon and axion/ALP.

We have included the contributions from the diagrams with $S$ coupling to both nucleons and the pion mediators, as shown in Fig.~\ref{fig:diagram}. We find that  to leading order in $m_S^2/m_N E_S$, the contributions from the $S-N-N$ diagrams cancel out with each other. As a result, the $S-N-N$ diagram contributions are suppressed by $(m_S/E_S)^4$. For light scalar with $m_S \lesssim 10$ MeV, the $S$-induced energy loss is therefore dominated by the diagrams with $S$ coupling to the pion mediators, as shown in Fig.~\ref{fig:comparison}.
%The separate contributions to energy loss rate from the diagrams with $S$ coupling to nucleons, pions and the cross terms are presented in Fig.~\ref{fig:comparison}.
In order to get the final supernova limits, we have calculated the decay and re-absorption rates of $S$, and find that the re-absorption via inverse nucleon bremsstrahlung $N N S \to N N$ plays a crucial role in the energy loss mechanism. The resultant dependence of the average mean free path on the scalar mass and mixing angle is shown in Fig.~\ref{fig:mfp}. With  the production, decay and re-absorption of $S$ properly taken into account, the supernova luminosity limit on $m_S$ and $\sin\theta$ is presented in Fig.~\ref{fig:limit}. Complementarity with the existing and future laboratory constraints is demonstrated in Fig.~\ref{fig:complementary}.

As pointed out in Refs.~\cite{Kazanas:2014mca, DeRocco:2019njg} (see also Ref.~\cite{Sung:2019xie}) for the case of dark photon, new limits on the properties of $S$ can arise from  the decay of $S$ inside the mantle of stars which can blow away the outer layers of the stellar material and generate intense light emission. This is likely if the $S$ decay products are mainly $e^+ e^-$, which then will annihilate and generate MeV $\gamma$-rays.  We do not consider this here since  this is beyond the main scope of this paper. Moreover, our calculations in this paper is at the leading order of OPE.  The beyond OPE effects have been considered for neutrino~\cite{Bartl:2016iok} and axion~\cite{Chang:2018rso, Carenza:2019pxu} production in the supernova core, and might also be important for the scalar $S$.
%The beyond leading-order effects as well as possible limits from $S$-decay near the mantle will be pursued in a future publication.
In addition to supernova explosion, the light scalar $S$ might also significantly impact neutron star mergers, through either  cooling the mergers or heat transportation inside the stars, similar to the axion case~\cite{Harris:2020qim}. As can be seen from Fig.~\ref{fig:complementary}, there exists an allowed parameter space around $\sin\theta\simeq 10^{-5}-10^{-4}$ (depending on the scalar mass) where the light scalar can be trapped in the supernova or neutron star core (see also Fig.~\ref{fig:mfp}), which could in principle affect the merger dynamics in a nontrivial way. A proper investigation of these issues, taking into account the complications arising due to the degenerate nuclear matter, nuclear equations of state and general relativity effects in a merger environment, is well beyond the scope of this work and will be pursued in a future communication.

\section*{Acknowledgments}

We are grateful to Mark Alford, Daniel Egana-Ugrinovic, Jean-Fran\c{c}ois Fortin, Steven Harris, Sam Homiller, Gordan Krnjaic, Patrick Meade, Shmuel Nussinov and Kuver Sinha for useful discussions and correspondence on the supernova limits on the CP-even scalar. Special thanks are due to Gordan Krnjaic for encouraging us to take up this project, to Steven Harris for helpful comments on the manuscript and for pointing out two errors in our codes for the calculation of mean free path, and to Kevin Kelly for a discussion on the DUNE near-detector sensitivity to light scalars and for providing us the data points for the DUNE sensitivity curve in Fig.~\ref{fig:complementary}.
The work of B.D. and Y.Z. is supported in part by the US Department of Energy under Grant No.  DE-SC0017987 and in part by the McDonnell Center for the Space Sciences. The work of R.N.M. is supported by the US National Science Foundation grant no. PHY-1914631.

\appendix

\section{Details of the emission rate calculation}
\label{sec:appendix}

\subsection{The amplitude}
\label{sec:amplitude}

The master formula of the emission rate $Q$ is given in Eq.~(\ref{eqn:rate}), and the corresponding Feynman diagrams are shown in Fig.~\ref{fig:diagram}. For the sake of concreteness let us denote the momentum of the nucleon $N_i$ in Fig.~\ref{fig:diagram} as $p_i$ ($i = 1,\, 2,\, 3,\, 4$), and the momentum of $S$ as $k_S$, the momenta exchanged in the $t$ and $u$-channels are then respectively $k \equiv p_2-p_4$ and $l \equiv p_2-p_3$,

Let us start with the proton bremsstrahlung process $p + p \to p + p + S$. The effective $p-p-\pi$ coupling is $(2m_N/m_\pi)f_{pp} \pi^0 \bar{p} i\gamma_5 p$ with $f_{pp} \simeq 1$. The amplitude for the diagram $(a)$ is
\begin{eqnarray}
\label{eqn:Ma}
{\cal M}_a & \ = \ &
\left[ \left( \frac{2m_N}{m_\pi} \right)^2
\frac{f_{pp}^2 (\sin\theta g_{hNN})}{k^2-m_\pi^2} \right]
\frac{1}{(p_3+k_S)^2 - m_N^2} \nonumber \\
&& \times \bar{u} (p_4) \gamma_5 u (p_2)
\bar{u} (p_3) ( \slashed{p}_3 + \slashed{k}_S + m_N ) \gamma_5 u (p_1)
\,.
\end{eqnarray}
The factors in the bracket are common to all the four diagrams $(a)$, $(b)$, $(c)$ and $(d)$ in the $t$-channel. As the nucleons are non-relativistic in the supernova core, we can use the approximation that $p_i \cdot k_S \simeq m_N E_S$ which is much larger than the other term in the propagator, i.e. $m_N E_S \gg m_S^2$. Then the second propagator in Eq.~(\ref{eqn:Ma}) can be simplified to be
\begin{eqnarray}
\label{eqn:LO}
\frac{1}{(p_i \pm k_S)^2 - m_N^2} \ \simeq \
\frac{1}{\pm 2 m_N E_S} \,.
\end{eqnarray}
Here we have included also the corresponding propagators for the diagrams $(b)$, $(c)$ and $(d)$, with the $+$ sign for the diagrams $(a)$ and $(b)$ with $i=3,\,4$ and the $-$ sign for $(c)$ and $(d)$ with $i=1,\,2$.
%For the diagram (b), this propagator is the same after the approximation. However, for (c) and (d),  they are
%\begin{eqnarray}
%\label{eqn:LO2}
%\frac{1}{(p_i-k_S)^2 - m_N^2} \ \simeq \
%-\frac{1}{2 m_N E_S} \quad {\rm with} \quad i = 1,\,2 \,.
%\end{eqnarray}
It is easy to obtain the spinor parts for the diagrams $(b)$, $(c)$ and $(d)$, which are respectively
\begin{eqnarray}
\label{eqn:pipj}
&& \bar{u} (p_3) \gamma_5 u (p_1)
\bar{u} (p_4) ( \slashed{p}_4 + \slashed{k}_S + m_N ) \gamma_5 u (p_2) \nonumber \,, \\
&& \bar{u} (p_4) \gamma_5 u (p_2)
\bar{u} (p_3) \gamma_5 ( \slashed{p}_1 - \slashed{k}_S + m_N ) u (p_1) \nonumber \,, \\
&& \bar{u} (p_3) \gamma_5 u (p_1)
\bar{u} (p_4) \gamma_5 ( \slashed{p}_2 - \slashed{k}_S + m_N )  u (p_2) \,.
\end{eqnarray}
Using the kinematic relations~\cite{Dent:2012mx}
\begin{eqnarray}
p_1\cdot p_2 & \ = \ &
  m_N^2 -\frac{1}{2} m_S^2 -\frac{1}{2} k^2 -\frac{1}{2} l^2
  -( k\cdot l) + (k\cdot k_S) + (l\cdot k_S) \nonumber \\
p_1\cdot p_3 &=& m_N^2 - \frac{1}{2} m_S^2 -\frac{1}{2} k^2 + (k\cdot k_S) \nonumber \, ,   \\
p_1\cdot p_4 &=& m_N^2 - \frac{1}{2} m_S^2 -\frac{1}{2} l^2 +  (l\cdot k_S) \nonumber \, ,  \\
p_2\cdot p_3 &=& m_N^2-\frac{1}{2} l^2 \nonumber  \, , \\
p_2\cdot p_4 &=& m_N^2-\frac{1}{2} k^2 \nonumber \, , \\
p_3\cdot p_4 &=&  m_N^2 -\frac{1}{2} k^2 -\frac{1}{2} l^2 + (k\cdot l) \, ,
\end{eqnarray}
it is easy to prove that with the approximation in Eq.~(\ref{eqn:LO}) the four amplitudes ${\cal M}_{a,b,c,d}$ cancel out with each other [see Eq.~(\ref{eqn:Mabcd})], as at the leading order in Eq.~(\ref{eqn:pipj}) $p_i \cdot p_j \simeq m_N^2$ with $i,\,j = 1,\,2,\,3,\,4$. To obtain a non-vanishing result, we expand the propagators to the next-to-leading order in $m_S^2/m_N E_S$, in the form of
\begin{eqnarray}
\frac{1}{(p_i \pm k_S)^2 - m_N^2} \ \simeq \
\frac{1}{\pm 2 m_N E_S + m_S^2}  \ \simeq \
\frac{1}{\pm 2 m_N E_S}
\left[ 1 \mp \frac{m_S^2}{2m_N E_S} \right]\,,
\end{eqnarray}
where the $+$ and $-$ signs are again respectively for $i=3,\,4$ and $1,\,2$. As a result of the opposite sign in the bracket for the diagrams $(a)$, $(b)$ and $(c)$, $(d)$, the sum of amplitudes are non-vanishing only at the next-to-leading order in $m_S^2/m_N E_S$.  The contributions of the diagrams $(a)$, $(b)$, $(c)$ and $(d)$ to the production rate will be suppressed by the ratio of $(m_S/E_S)^4$ in the limit of small $m_S$ [see Eq.~(\ref{eqn:Itot})].

The amplitude for the $t$-channel diagram $(e)$ is
\begin{eqnarray}
{\cal M}_e & \ = \ &
\left( \frac{2m_N}{m_\pi} \right)^2
\frac{f_{pp}^2 \sin\theta {\cal A}_\pi}{k^2-m_\pi^2}
\frac{1}{(k-k_S)^2 - m_\pi^2} \,
\bar{u} (p_4) \gamma_5 u (p_2)
\bar{u} (p_3) \gamma_5 u (p_1)
\,.
\end{eqnarray}
It is expected that $|{\bf k_S}| \sim T$, and $|{\bf k}|^2 \sim m_N T$. We can  then neglect the momentum $k_S$ in the second propagator. The subsequent calculation of ${\cal M}_e$ is trivial.

Evaluations of the $u$-channel diagrams $(a')$, $(b')$, $(c')$, $(d')$ and $(e')$ and the $t$ and $u$-channel cross terms are quite similar (note that there is an overall relative minus sign for the $u$-channel diagrams). With the approximations $k^2 \simeq -|{\bf k}|^2$, $l^2 \simeq -|{\bf l}|^2$, $(k \cdot l) \simeq -({\bf k} \cdot {\bf l})$ and ${\bf k}^2 ,\, {\bf l}^2 ,\, ({\bf k} \cdot {\bf l}) \gg m_S^2,\, E_S^2$, the squared amplitude for the $pp$ process is proportional to
%Combing the five diagrams in the $t$-channels and the five in the $u$-channel, the total amplitude reads
%\begin{eqnarray}
%{\cal M}^{(pp)} \ = \ \sum_{i=a,b,c,d} {\cal M}_i + {\cal M}_e -
%\sum_{i=a',b',c',d'} {\cal M}_{i^\prime} - {\cal M}_{e'} \,,
%\end{eqnarray}
%which can be functions of , $v_{\rm EW}$, the masses $m_N$, $m_\pi$ and $m_S$, the coupling constants and the mixing angle $\sin\theta$, i.e.
\begin{eqnarray}
\label{eqn:Msq}
\frac{y_{hNN}^2 m_S^4}{E_S^4} I_A^{(pp)}
+ \frac{m_N^2}{81 v_{\rm EW}^2} I_B^{(pp)}
+ \frac{2}{9} \frac{y_{hNN}m_S^2}{E_S^2} \frac{m_N}{v_{\rm EW}} I_C^{(pp)} \,,
\end{eqnarray}
where the dimensionless functions of the momenta are respectively
\begin{eqnarray}
I_A^{(pp)} & \  = \ & \frac{{\bf k}^4}{({\bf k}^2 + m_\pi^2)^2} +
\frac{{\bf l}^4}{({\bf l}^2 + m_\pi^2)^2} -
\frac{{\bf k}^2 {\bf l}^2 - 2 ({\bf k} \cdot {\bf l})^2}{({\bf k}^2 + m_\pi^2)({\bf l}^2 + m_\pi^2)} \,, \\
%%%%%%%%%%%%%%%%%%%%%%%%%%%%%%%%%%%%%%%%%%%%%%%%%%%%%%%%%%%%%%%%%%
I_B^{(pp)} & \ = \ &
\left( m_S^2 + \frac{11}{2} m_\pi^2 \right)^2
\left[  \frac{{\bf k}^4}{({\bf k}^2 + m_\pi^2)^4} +
\frac{{\bf l}^4}{({\bf l}^2 + m_\pi^2)^4} -
\frac{{\bf k}^2 {\bf l}^2 - 2 ({\bf k} \cdot {\bf l})^2}{({\bf k}^2 + m_\pi^2)^2({\bf l}^2 + m_\pi^2)^2} \right] \,, \nonumber \\ && \\
%%%%%%%%%%%%%%%%%%%%%%%%%%%%%%%%%%%%%%%%%%%%%%%%%%%%%%%%%%%%%%%%%%
I_C^{(pp)} & \ = \ &
\left( m_S^2 + \frac{11}{2} m_\pi^2 \right)
\left[  \frac{{\bf k}^4}{({\bf k}^2 + m_\pi^2)^3} +
\frac{{\bf l}^4}{({\bf l}^2 + m_\pi^2)^3}
-\frac12 \frac{{\bf k}^2 {\bf l}^2 - 2 ({\bf k} \cdot {\bf l})^2}{({\bf k}^2 + m_\pi^2)({\bf l}^2 + m_\pi^2)^2} \right. \nonumber \\
&& \left. -\frac12 \frac{{\bf k}^2 {\bf l}^2 - 2 ({\bf k} \cdot {\bf l})^2}{({\bf k}^2 + m_\pi^2)^2({\bf l}^2 + m_\pi^2)}\right] \,.
\end{eqnarray}
Here the $I_{A,\,B}^{(pp)}$ terms are respectively the contributions from the diagrams with the scalar $S$ coupling to the protons and pions, and  $I_{C}^{(pp)}$ are from the cross terms.

Let us now move on to the $nn$ and $np$-channel production of $S$. Isospin invariance requires that $f_{nn} = - f_{pp}$ and $f_{np} = - \sqrt2 f_{pp}$. This implies that the $nn$ contribution is the same as that for the $pp$ process, i.e.
\begin{eqnarray}
I_{A,\,B,\,C}^{(nn)} \ = \  I_{A,\,B,\,C}^{(pp)} \,.
\end{eqnarray}
For the $np$ processes, there will be an extra factor of $(-\sqrt2)^2 = 2$ for the amplitudes of $u$-channel diagrams. When amplitudes are squared, there will be an extra factor of 4 for the $u$-channel diagrams, and a factor of $1\times (-1) \times (-\sqrt2)^2 = -2$ for cross terms of $t$ and $u$-channel diagrams. As a result, we get
\begin{eqnarray}
I_A^{(np)} & \  = \ & \frac{{\bf k}^4}{({\bf k}^2 + m_\pi^2)^2} +
\frac{4{\bf l}^4}{({\bf l}^2 + m_\pi^2)^2} +
\frac{2 \left[{\bf k}^2 {\bf l}^2 - 2 ({\bf k} \cdot {\bf l})^2 \right]}{({\bf k}^2 + m_\pi^2)({\bf l}^2 + m_\pi^2)} \,, \\
%%%%%%%%%%%%%%%%%%%%%%%%%%%%%%%%%%%%%%%%%%%%%%%%%%%%%%%%%%%%%%%%%%
I_B^{(np)} & \ = \ &
\left( m_S^2 + \frac{11}{2} m_\pi^2 \right)^2
\left[  \frac{{\bf k}^4}{({\bf k}^2 + m_\pi^2)^4} +
\frac{4{\bf l}^4}{({\bf l}^2 + m_\pi^2)^4} +
\frac{ 2\left[{\bf k}^2 {\bf l}^2 - 2 ({\bf k} \cdot {\bf l})^2\right]}{({\bf k}^2 + m_\pi^2)^2({\bf l}^2 + m_\pi^2)^2} \right] \,, \nonumber \\ && \\
%%%%%%%%%%%%%%%%%%%%%%%%%%%%%%%%%%%%%%%%%%%%%%%%%%%%%%%%%%%%%%%%%%
I_C^{(np)} & \ = \ &
\left( m_S^2 + \frac{11}{2} m_\pi^2 \right)
\left[ \frac{{\bf k}^4}{({\bf k}^2 + m_\pi^2)^3} +
\frac{4{\bf l}^4}{({\bf l}^2 + m_\pi^2)^3}
+ \frac{{\bf k}^2 {\bf l}^2 - 2 ({\bf k} \cdot {\bf l})^2}{({\bf k}^2 + m_\pi^2)({\bf l}^2 + m_\pi^2)^2} \right. \nonumber \\
&& \left. + \frac{{\bf k}^2 {\bf l}^2 - 2 ({\bf k} \cdot {\bf l})^2}{({\bf k}^2 + m_\pi^2)^2({\bf l}^2 + m_\pi^2)}\right] \,.
\end{eqnarray}

Summing up the $pp$, $nn$ and $np$ channels, we get the full squared amplitude expression:
\begin{eqnarray}
\label{eqn:Msq}
\sum_{\rm spins} \left| {\cal M}^{} \right|^2 \ & \simeq & \
\frac{256 \pi^2 \alpha_\pi^2 f_{pp}^4 \sin^2\theta}{m_N^2}
\left[ \frac{y_{hNN}^2 m_S^4}{E_S^4} I_A^{}
+ \frac{m_N^2}{81 v_{\rm EW}^2} I_B^{} \right. \nonumber \\
&& \left. \quad
+ \frac{2}{9} \frac{y_{hNN}m_S^2}{E_S^2} \frac{m_N}{v_{\rm EW}} I_C^{} \right] \,,
\end{eqnarray}
where  %$\alpha_\pi \equiv (2m_N/m_\pi)^2/4\pi \simeq 15$, and
\begin{eqnarray}
\label{eqn:IABC}
I_{A,\,B,\,C} \ & = & \ I_{A,\,B,\,C}^{(pp)} +
I_{A,\,B,\,C}^{(nn)} + 4I_{A,\,B,\,C}^{(np)} \,.
\end{eqnarray}
%Replacing the quantities $I_{A,\,B,\,C}^{(pp)}$ by $I_{A,\,B,\,C}$, we finally obtain the squared amplitude $\sum_{\rm spins} \left| {\cal M}^{} \right|^2$.
As in Eq.~(\ref{eqn:Iabc}), the factor of $4$ for the $np$ channel is for non-identical particles in the initial and final states.
%For the $pp$ and $nn$ channels, there is a factor of $1/2 \times 1/2  = 1/4$ for identical particles in both the initial and final states, thus we have an extra factor of $4$ for the $np$ term in Eq.~(\ref{eqn:IABC}).

\subsection{The emission rate}
\label{sec:rate}

To simplify the fifteen-dimensional phase space integral in Eq.~(\ref{eqn:rate}) for the emission rate, we define the following dimensionless variables~\cite{Giannotti:2005tn, Dent:2012mx}:
\begin{align}
\label{eqn:uvxyqz}
& u \ \equiv \ \frac{{\bf p}_i^2}{m_N T} \,, \qquad
 v \ \equiv \ \frac{{\bf p}_f^2}{m_N T} \,, \qquad
 x \ \equiv \ \frac{E_S}{T} \,,  \qquad
 y \ \equiv \ \frac{m_{\pi}^2}{m_N T} \,, \nonumber \\
&  r \ \equiv \ \frac{T}{m_N} \,, \qquad
 q \ \equiv \ \frac{m_S}{T} \,, \qquad
 z \ \equiv \ \cos(\theta_{if}) \,,
\end{align}
with
\begin{align}
& \quad {\bf p}_1  \ \equiv \  {\bf P} + {\bf p}_i \,, \quad
{\bf p}_2 \ \equiv \ {\bf P} - {\bf p}_i \,, \quad %\nonumber \\
{\bf p}_3  \ \equiv \  {\bf P} + {\bf p}_f \,, \quad
{\bf p}_4 \ \equiv \ {\bf P} - {\bf p}_f \,.
\end{align}
We have taken the approximation that ${\bf p}_i \gg {\bf k_S}$ such that ${\bf p}_3 + {\bf p}_4 \simeq {\bf p}_1 + {\bf p}_2 = {\bf P}$. In Eq.~(\ref{eqn:uvxyqz}) $z$ is defined as the angle $\theta_{if}$ between ${\bf p}_i$ and ${\bf p}_f$. Then it is trivial to obtain the following relations:
\begin{eqnarray}
{\bf k}^2 & \ = \ &  (u+v-2z\sqrt{uv}) m_N T \ \equiv \ c_k m_N T  \,, \\
{\bf k}^2 + m_\pi^2 & \ = \ &  (u+v+y-2z\sqrt{uv}) m_N T \ \equiv \ c_{k\pi} m_N T  \,, \\
{\bf l}^2 & \ = \ &  (u+v+2z\sqrt{uv}) m_N T \ \equiv \ c_l m_N T \,, \\
{\bf l}^2 + m_\pi^2 & \ = \ &  (u+v+y+2z\sqrt{uv}) m_N T \ \equiv \ c_{l\pi} m_N T  \,, \\
{\bf k}^2 {\bf l}^2 - 2 ({\bf k}\cdot {\bf l})^2 & \ = \ &
 \left[ u^2 + v^2 + 2 u v (-3 + 2 z^2) \right] m_N^2 T^2  \ \equiv \ (c_{kl} m_N T)^2\,.
\end{eqnarray}
It is now straightforward to arrive at the final expression for the emission rate $Q$ in Eq.~(\ref{eqn:rate2}), %with the dimensionless functions defined as:
%and the emission rat (\ref{eqn:rate}) can be simplified to be
%\begin{eqnarray}
%\label{eqn:rate2}
%Q & \ = \ & \frac{n_B^2 \alpha_\pi^2 f_{pp}^4 T^{7/2} \sin^2\theta}{8 \pi^{3/2} m_N^{9/2}} \int {\rm d}u {\rm d}v {\rm d}z {\rm d}x
%\sqrt{uv} e^{-u} x \sqrt{x^2 - q^2} \delta (u-v-x) R_{\rm decay} R_{\rm abs} {\cal I}_{\rm tot} \,, \nonumber \\
%\end{eqnarray}
%with
%\begin{eqnarray}
%\label{eqn:Itot}
%{\cal I}_{\rm tot} \ = \
%y_{hNN}^2 \left( \frac{q}{x} \right)^4 {\cal I}_A +
%\frac{1}{81} \left( \frac{m_N}{v_{\rm EW}} \right)^2 {\cal I}_B +
%\frac{2}{9} y_{hNN} \left( \frac{q}{x} \right)^2
%\left( \frac{m_N}{v_{\rm EW}} \right) {\cal I}_C \,,
%\end{eqnarray}
with the dimensionless components for the $pp$, $nn$ and $np$ processes respectively: %(the $nn$ process contribution is the same as the $pp$ process):
\begin{eqnarray}
\label{eqn:IApp}
{\cal I}_{A}^{(pp)} & \ = \ &  {\cal I}_{A}^{(nn)}  \ = \
\frac{c_k^2}{c_{k\pi}^2} + \frac{c_l^2}{c_{l\pi}^2} - \frac{c_{kl}^2}{c_{k\pi}c_{l\pi}} \,, \\
%%%%%%%%%%%%%%%%%%%%%%%%%%%%%%%%%%%%%%%%%%%%%%%%%%%%%%%%%%%%%%%%%%
\label{eqn:IBpp}
{\cal I}_{B}^{(pp)} & \ = \ &  {\cal I}_{B}^{(nn)}  \ = \
\left( q^2 r + \frac{11}{2} y \right)^2
\left[ \frac{c_k^2}{c_{k\pi}^4} + \frac{c_l^2}{c_{l\pi}^4} - \frac{c_{kl}^2}{c_{k\pi}^2c_{l\pi}^2} \right] \,, \\
%%%%%%%%%%%%%%%%%%%%%%%%%%%%%%%%%%%%%%%%%%%%%%%%%%%%%%%%%%%%%%%%%%
\label{eqn:ICpp}
{\cal I}_{C}^{(pp)} & \ = \ &  {\cal I}_{C}^{(nn)}  \ = \
\left( q^2r + \frac{11}{2} y \right)
\left[ \frac{c_k^2}{c_{k\pi}^3} + \frac{c_l^2}{c_{l\pi}^3} - \frac{c_{kl}^2}{2c_{k\pi}c_{l\pi}^2} - \frac{c_{kl}^2}{2c_{k\pi}^2c_{l\pi}} \right] \,, \\
%%%%%%%%%%%%%%%%%%%%%%%%%%%%%%%%%%%%%%%%%%%%%%%%%%%%%%%%%%%%%%%%%%
%%%%%%%%%%%%%%%%%%%%%%%%%%%%%%%%%%%%%%%%%%%%%%%%%%%%%%%%%%%%%%%%%%
\label{eqn:IAnp}
{\cal I}_{A}^{(np)} & \ = \ &
\frac{c_k^2}{c_{k\pi}^2} + \frac{4c_l^2}{c_{l\pi}^2} + \frac{2c_{kl}^2}{c_{k\pi}c_{l\pi}} \,, \\
%%%%%%%%%%%%%%%%%%%%%%%%%%%%%%%%%%%%%%%%%%%%%%%%%%%%%%%%%%%%%%%%%%
\label{eqn:IBnp}
{\cal I}_{B}^{(np)} & \ = \ &
\left( q^2 r + \frac{11}{2} y \right)^2
\left[ \frac{c_k^2}{c_{k\pi}^4} + \frac{4c_l^2}{c_{l\pi}^4} + \frac{2c_{kl}^2}{c_{k\pi}^2c_{l\pi}^2} \right] \,, \\
%%%%%%%%%%%%%%%%%%%%%%%%%%%%%%%%%%%%%%%%%%%%%%%%%%%%%%%%%%%%%%%%%%
\label{eqn:ICnp}
{\cal I}_{C}^{(np)} & \ = \ &
\left( q^2 r + \frac{11}{2} y \right)
\left[ \frac{c_k^2}{c_{k\pi}^3} + \frac{4c_l^2}{c_{l\pi}^3} + \frac{c_{kl}^2}{c_{k\pi}c_{l\pi}^2} + \frac{c_{kl}^2}{c_{k\pi}^2c_{l\pi}} \right] \,.
\end{eqnarray}

Simplification of the twelve-dimensional integration in Eq.~(\ref{eqn:mfp}) for the inverse MFP $\lambda^{-1}$ is quite similar, which leads to the formula in Eq.~(\ref{eqn:mfp2}).

\bibliographystyle{JHEP}
\bibliography{ref}

\end{document}